# Coupling component systems towards systems of systems

Frédéric Autran, Jean-Philippe Auzelle, Denise Cattan, Jean-Luc Garnier, Dominique Luzeaux, Frédérique Mayer, Marc Peyrichon, Jean-René Ruault

Association Française d'Ingénierie Système

Parc Club Orsay, 32, Rue Jean Rostand

91893 ORSAY Cedex, France

e-mail afis@afis.fr

**Abstract.** Systems of systems (SoS) are a hot topic in our "fully connected global world". Our aim is not to provide another definition of what SoS are, but rather to focus on the adequacy of reusing standard system architecting techniques within this approach in order to improve performance, fault detection and safety issues in large-scale coupled systems that definitely qualify as SoS, whatever the definition is. A key issue will be to secure the availability of the services provided by the SoS despite the evolution of the various systems composing the SoS. We will also tackle contracting issues and responsibility transfers, as they should be addressed to ensure the expected behavior of the SoS whilst the various independently contracted systems evolve asynchronously.

## Introduction

Systems of systems (SoS) can be defined loosely as a combination of systems in order to fulfill some kind of capability, with the additional fact that the composing systems should have operational and managerial independence. We will not delve into the current debate of looking for the appropriate definition, since our aim is to start from a real-world generic example and address concrete issues, which can be used later to feed the current debate.

Henceforth we will deal with several systems that already provide services to their customers/users, and that are coupled with some new structure – which we dare to call a SoS – that provides new (emergent) services to the customers/users. The coupling creates added value on the one hand, as new services are available, but it increases the appearance of failure modes within the whole chain value on the other hand.

This situation is more and more likely to occur, as multinational cooperation is necessary to efficiently handle either major crisis, prevent disasters or just simply make a better usage of the natural resources. In this way, system of systems engineering can be seen as a tool to serve the Earth!

We will show that a straightforward extension of the standard functional dependence coupling matrix can be used to provide adequate answers.

## Definitions: coupling matrix and system of systems

A key driver to understanding the non-triviality of the current debate on SoS is that, following the generally accepted definition, a system delivers products and/or services. Hence the combination of systems gives birth to a tangle of products and services, which justifies the search for an encompassing concept but adds to the general confusion. The merging of tangible and immaterial value creating entities actually contributes to the complexity of the resulting structure.

Among the popular definitions of SoS, Mark Maier's definition [MAI98b] underlines the following properties:
- *Operational independence of the elements: if the SoS is disassembled into its component systems the component systems must be able to operate independently in an efficient way. The SoS is composed of systems which are independent and useful in their own right.*
- *Managerial independence of the elements: the component systems not only can operate independently, they do operate independently. The component systems are acquired and integrated separately but maintain a continuing operational existence independent of the SoS.*
- *Evolutionary development: the SoS does not appear fully formed. Its development and existence is evolutionary with functions and purposes added, removed, and modified with experience.*
- *Emergent behavior: the system performs functions and carries out purposes that do not reside in any component system. These behaviors are emergent properties of the entire SoS and cannot be claimed by any component system. The principal purposes of the SoS are fulfilled by these behaviors.*
- *Geographic distribution: the geographic extent of the component systems is large. Large is a nebulous and relative concept as communication capabilities increase, but at a minimum it means that the components can readily exchange only information and not substantial quantities of mass or energy.*

Our approach to managing the creation of value obtained through the combination of systems is to adopt a service-oriented picture, and adapt current "product-driven" system engineering tools and use them as "service-driven" SoS engineering tools. Indeed when combining systems for which products are exchanged, consumed, or transformed, the resulting added value is a priori not greater than the sum of all added values of the component systems. However when considering services (which are immaterial, can be composed, and have added-value for the consumer and the provider in a predefined context of use – cf. ISO/CEI20000) the story changes. Collaboration between service-providing systems allows realizing higher-level services which contribute to the added value of the target SoS. Henceforth in the sequel, we will mainly discuss services and relegate products to the background.

The engineering tool used extensively in this paper is the $N^2$ dependence coupling matrix [MEI98, MEI02]. It is used to combine the components into sub-systems, including the communication means. One aims at obtaining sub-systems with a strong/high internal cohesion and a loose external coupling.

In our context of SoS, the components are systems. The connections/links/interfaces vehicle combinations of products and services. Such combinations can be either sequential or more complex (parallel, braiding…) combinations. By identifying dependence and collaboration between these service-providing systems, one wants to enhance and preserve the added value of the target SoS and to manage the configuration of the latter.

Figure 1 illustrates the former notions with an example applied to functional flows. On the left side the coupling between four systems (S1, S2, S3 and S4) is represented by a flow diagram. The right side shows the corresponding coupling matrix.

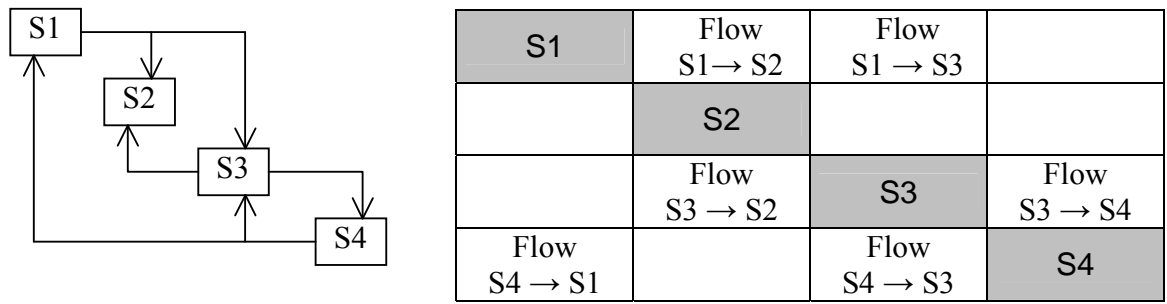

Figure 1. Coupled systems and the resulting coupling matrix.

The systems building up the SoS lie on the diagonal of the coupling matrix, whereas the flow exchanged between a source component i and a target component j of the SoS lies on the corresponding (i,j) cell (the row corresponds to the source component while the column is the target component).

As an additional feature, we can associate to each cell various information, such as the necessary resources in order to fulfill the relevant service, or critical parameters/constraints to take into account for the safety or the nominal functioning of the systems. Actually this additional information can be organized so as to yield various architecture views of the SoS, similar to what is commonly done by system architects.

Furthermore, we will detail how it is possible to use this matrix to depict dependencies other than flows such as physical interfaces, contractual management or legal rules.

One of the advantages of this representation is to yield an easy way to read emerging functions and services: if there is a path – or a set of paths – leading from a source cell (s,s) to a target cell (t,t), i.e. a chain of dependencies $[(s,s),(s,i_1),(i_1,i_2),…,(i_n,t),(t,t)]$, then the combination (sequential, parallel, etc.) of all that services defines a new service, that can be denoted as emerging since it was not foreseen initially. It can be defined informally in natural language, as can be seen from the example below, but more interestingly it can be defined formally when one looks at the associated resources and when one knows how to compose formally the services.

We will not detail any formal technique in this paper that can be used to define compositionally the emerging services, but they are similar to what is used in process calculus (e.g. Milner's pi-calculus [MIL99]) in mobile communication theory. Furthermore, formal verification techniques can be defined that rely on particular logics, like linear logic (cf. Girard's and Lafont's work on linear logic proof nets [Abrusci 95]), that take resource consumption into account. This shows how the seemingly trivial representation above can be used extensively throughout the design and verification processes within the engineering of the SoS.

# Context and case study

## *Context*

We will use below a case study to explain these concepts and to elaborate the functional dependence coupling matrix. This case study is based upon a fire emergency situation, extracted from the Press file "Protect forests against fires" [DDSC 07].

The scenario deals with a forest fire in a mountainous area. Various fire fighting devices can be used, depending on the geographical location, dwelling proximity, weather and weather evolution, and availability of aerial crews and ground squads.

Integration of systems provides new capabilities to the whole SoS. These new capabilities which lead to improved global performance are:
- optimization of firefighting resources,
- reduction of the response time to the initial outbreak which prevents fire from spreading to a conflagration out of control and thus saves lives, wildlife and natural resources.

During the "warning" stage of firefighting, the departmental operation center continuously updates the fire risk assessment for the concerned areas. The departmental operation center prepositions resources according to the fire risk. Early warning is raised as soon as aerial watch and ground watch detect any fire start.

At this point of time, different scenarios can occur. Let us consider the case where an unpredictable expansion of the fire is caused by exceptional weather conditions and requires additional forces. Those additional forces are coming from foreign firefighting forces.

Our assumption is that both national command chains are deployed. Different possible scenarios can be considered for the coordination that can be qualified using the OIM (Organizational Interoperability Maturity model) scale, ranging from 0 (Independent) to 4 (Unified) [CLARK99]. In the case of our illustration of highly interacting systems, we want to achieve level 3 (Integrated) that means that:
- *the integrated level of organizational interoperability is one where there are shared value systems and shared goals,*
- *a common understanding and a preparedness to interoperate, for example, detailed doctrine is in place and there is significant experience in using it.*
- *the frameworks are in place and practiced however there are still residual attachments to a home organization.*

**Disclaimer**: The following is only a case illustration of coupling component systems towards systems of systems. This picture is intentionally simplified for our purpose.

## *Description of the fire emergency System of Systems*

The **emergency operation command center** (1 – Figure 2) is an ad hoc organization that controls the whole fire means and:
[1.1]: establishes the top priority to mobile headquarter,
[1.2]: requests additional means and necessary actions to the departmental operation center,
[1.3]: informs mobile head quarter about situational evolution,
[1.4]: provides the departmental operation center with information regarding operation in progress as well as evolution of situation.

The **mobile headquarter** (2 - Figure 2) is positioned near the disaster and moves regularly on the ground at a security range from the fire. The mobile HQ is the nerve center of the system:

[2.1]: reports the fire status to the emergency operation command center,
[2.2]: decides tactic of fight to adopt by the ground firefighter squad,
[2.3]: requests resources to the departmental operation center,
[2.4]: coordinates (by voice and radio) both the water bombers and the groups of firefighters present on the ground (preventing them from aerial droppings).

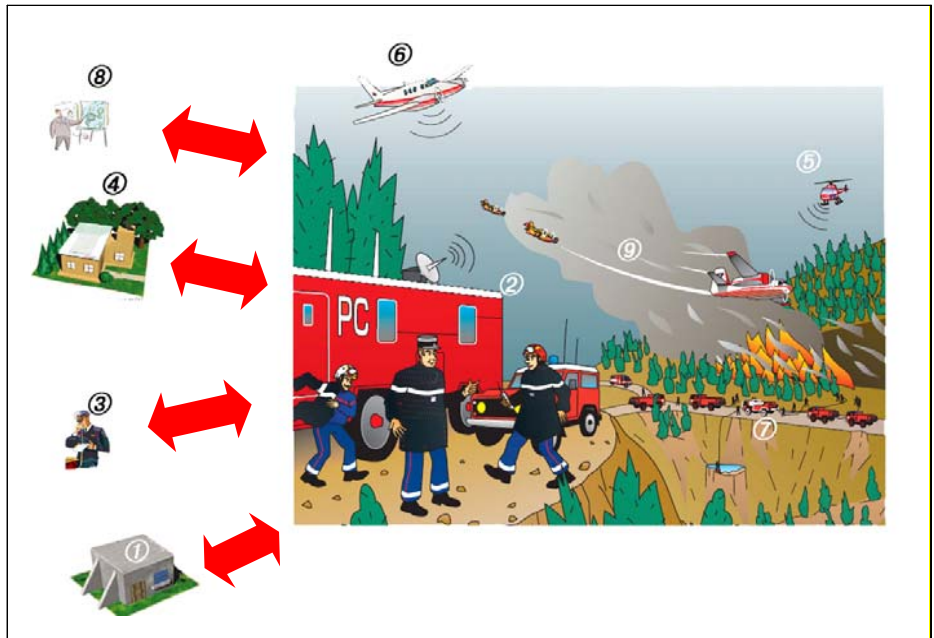

Figure 2. Operational view of the Emergency Firefighting System (EFS).

The **air officer** (3 – Figure 2):
[3.1]: provides assistance to the commander and the coordination plane regarding the air means,
[3.2]: controls the water bombers and assigns them their missions.

The **departmental operation center** (4 – Figure 2):
[4.1]: manages the local and national resources made available to the departmental service of fire and emergency,
[4.2]: delivers to the ground firefighter squads the evacuation order for the population,
[4.3]: orders engagement of additional means: firemen, foresters sappers, units soldiers and air means necessary to the execution of the operation, this is external interface outside the scope of this issue,
[4.4]: provides mobile HQ with all information regarding the actual resources, position of the active hearths and the fire line, localization of the "significant" points (dwellings, camp-site…).

The **command helicopter** (5 – Figure 2) hovering with the top of the flames:
[5.1]: delivers a global vision of the disaster to the mobile headquarter,
[5.2]: is used to mark the objectives which the water bomber planes must reach, to the mobile

headquarter.

The **coordination plane** (6 – Figure 2), watching at 1,500 feet above the fire area:
[6.1]: manages the coordination and the safety of the various planes in real time,
[6.2]: indicates to the water bombers pilots the obstacles, like high voltage lines, to carry out their droppings,
[6.3]: provides air officer with air traffic information.

The **ground firefighter squads** (7 – Figure 2):
[7.1]: report fire situation to mobile HQ.

The **weather team** (8 – Figure 2):
[8.1]: gathers data from national weather forecast services, the network of the automatic stations, the weather mobile cases, results of sample analysis, etc., this is external interface outside the scope of this issue,
[8.2]: provides the departmental operation center, the emergency operation command center, the air officer and the water bombers with weather data.

The **water bombers** (9 – Figure 2):
[9.1]: report the water-drops in real-time to the mobile headquarter,
[9.2]: report the environmental information to the air officer.

## Resulting coupling matrix

The previous description is easily translated into the following coupling matrix.
This matrix allows verifying loops of interactions between systems (e.g. servo control) and the source and pit of information. In the case of the coordination plane, it gets information from environment to elaborate and to distribute situation picture. It's a source, without system input. It is the same situation for the weather team.

Table 1: Coupling matrix for the EFS scenario.

| 1 | [1.1], [1.3] |  | [1.2] , [1.4] |  |  |  |  |  |
|---|---|---|---|---|---|---|---|---|
| [2.1] | 2 |  | [2.3] |  |  | [2.2] , [2.4] |  | [2.4] |
|  | [3.1] | 3 |  |  | [3.1] |  |  | [3.2] |
|  | [4.4] |  | 4 |  |  | [4.2] , [4.3] |  |  |
|  | [5.1] , [5.2] |  |  | 5 |  |  |  |  |
|  |  | [6.3] |  |  | 6 |  |  | [6.1] , [6.2] |
|  | [7.1] |  |  |  |  | 7 |  |  |
| [8.2] |  | [8.2] | [8.2] |  |  |  | 8 | [8.2] |
|  | [9.1] | [9.2] |  |  |  |  |  | 9 |

When some fire fighter resources coming from various different countries intervene together, the matrix becomes more complex. In this case, there are n² configurations, with n equal le number of countries intervening.

The following table shows three SoS intervening. The departmental operation center manages the local and national resources made available to the departmental service of fire and emergency. Mobile head quarters share information. Air officers exchange information about air traffic management. Ground firefighter squads coordinate themselves smoothly.

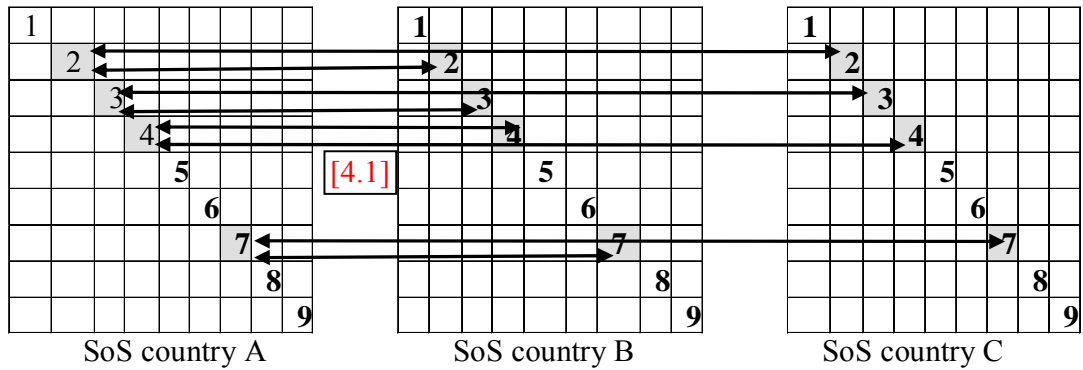

Figure 3. SoS interoperability in the case of multinational intervention.

Direct communication between ground forces requires that firefighters speak a common language. If it is so they can coordinate themselves, accurately, without problem. If they don't speak a common language (English for instance), it is necessary to define a liaison agent, interface between two groups. This liaison agent can insure both horizontal and vertical operability.

## Dependence matrix: compatibility and interoperability issues

Since the SoS consists of several interconnected systems which have been designed a priori without knowledge of each other, the various assumptions about the external world of each system may conflict, leading to compatibility problems (e.g. incompatible frequency plans of transmitters used in a water bomber and ground teams). These are special cases of *interoperability* issues, which are crucial to allow any service exchange between technical systems. Interoperability is not restricted to the existence of physical links between the systems. It occurs at various levels; for instance, NATO defines three levels of interoperability for military systems:

- Physical interoperability: a communication link must exist. This link can be wired or wireless, and is not necessarily IT-based, e.g. voice can be used as a communication medium.
- Procedural interoperability: a protocol and a syntactical form must be known and used for exchange.
- Operational interoperability: it refers to the activities related to the operation of a system in the context of other systems, e.g. doctrine governing the way the system is used. We differentiate the IT side of the operational interoperability (semantic interoperability between services) and the user side, i.e. how he understands information (sense-making and shared situational awareness) [EBR03], [SAS06], [WAR04].

An obvious solution for interoperability is to define an interface for each pair of communicating systems. However this can be achieved only for a small-scale system of systems, since for a large one, the number of interfaces necessarily leads to a very high cost, when feasible.

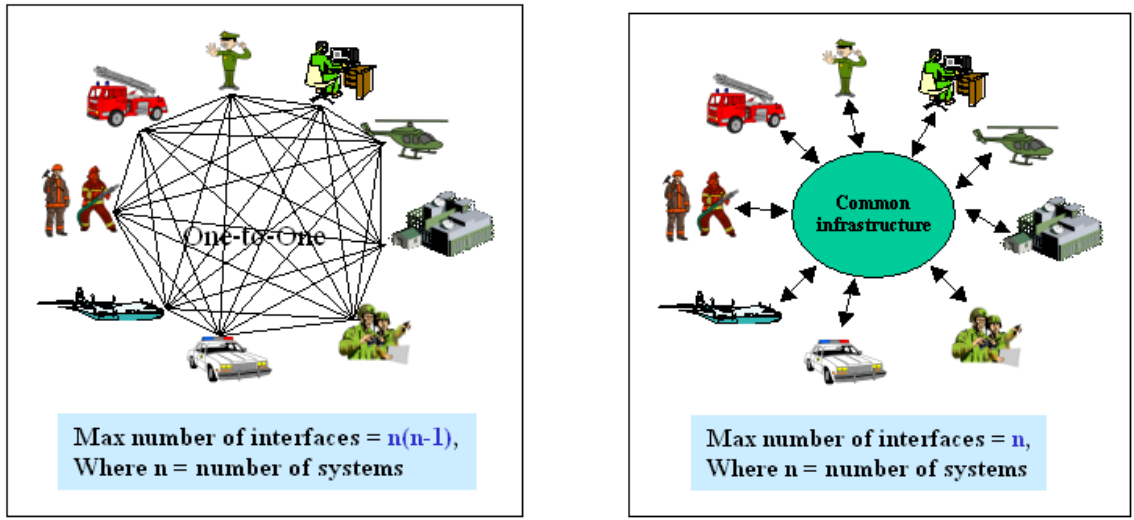

Figure 4. Common infrastructure decreases complexity.

Other solutions have to be put in place for interoperability, e.g.:
- Usage of a common technical infrastructure for physical interoperability: in that case the systems are no longer peer-to-peer linked, but each one is linked to the infrastructure.
- Usage of a common service-oriented infrastructure for procedural interoperability: this constitutes the current paradigm for information or IT-driven systems, and leads to service-oriented architectures. A "service repository" is expected to facilitate loosen the coupling between systems. It is empty when created and knows neither providers nor consumers. The service-providing systems access the repository to store the service definition in a neutral representation (location of the access point to invoke the service, service parameters, and quality of service). The service-consuming systems access the repository to get a service according to their need, and the invocation of the relevant service is then performed. This service repository plays the role of mediator and third party, and therefore enhances the security of the system by masking the service provider to the service consumer.
- Semantic issues for operational interoperability are more complex and not fully mature within the IT domain. They rely on the definition of common dictionaries (called "meta-data registry" in the US literature and "pivot model" in French) which are widely used for information systems and provide a public data model that allows communication between the different systems.

This multifaceted infrastructure which ensures general interoperability has to be included within the SoS as a new component with specific couplings with the relevant systems, and it has obviously its own life-cycle: in particular an adequate configuration management should be performed. Indeed a common infrastructure facilitates the architecture of the SoS at a given time, but not on the longer term: there is a necessary trade-off between immediate added facility (which increases short-term agility) and increased configuration management in time.

Coming back to our case study, the following table uses the coupling matrix and focuses on exchange and compatibility of services (which are mainly data in our case), indicating the versions of the exchanged services (service version on provider side, infrastructures version, service version on client side). This means that the system is currently running with the set of service versions. From this matrix it is straightforward to see the compliance requirements for a service: just look at the column to which the given service pertains. In the case of a system consuming services from several independent provider systems, the former must be compatible with all the interfaces of the latter through service version adaptation. For instance, the mobile HQ must receive information from quite all other elementary systems. Hence mobile HQ needs to have the necessary adapters to translate the incoming data into its proper data model.

Table 2: Services and information, exchange and compatibility between systems.

| 1 | [1.1] / (1.6/4.3/2.2) [1.3] / (3.2/5.2/2.0) | | [1.2], (./././.) [1.4], (./././.) | | | | | |
|---|---|---|---|---|---|---|---|---|
| [2.1], (./././.) | 2 | | [2.3], (./././.) | | | [2.2], (./././.) [2.4], (./././.) | | [2.4], (./././.) |
| | [3.1], (./././.) | 3 | | | [3.1], (./././.) | | | [3.2], (./././.) |
| | [4.4], (./././.) | | 4 | | | [4.2], (./././.) [4.3], (./././.) | | |
| | [5.1], (./././.) [5.2], (./././.) | | | 5 | | | | |
| | | [6.3], (./././.) | | | 6 | | | [6.1], (./././.) [6.2], (./././.) |
| | [7.1], (./././.) | | | | | 7 | | |
| [8.2], (./././.) | | [8.2], (./././.) | [8.2], (./././.) | | | | 8 | [8.2], (./././.) |
| | [9.1], (./././.) | [9.2], (./././.) | | | | | | 9 |

In above table, the first gray cell indicates the versions 1.6, 4.3 and 2.2 of the [1.1] exchange and versions 3.2, 5.2 and 2.0 of the [1.3] exchange, respectively for provider, infrastructure and client sides. Other cells of the table are not valued; the version numbers are replaced with dots.

## Managerial independence, system's owner and manager issues

While the coupling of systems provides new capabilities and services, managerial independence of elements of the SoS means that each system is managed independently, including the evolution of the provided services, or the updating of data flows and interfaces. Each system evolves apart, depending on its owner's or manager's goals, needs and means. On the other hand, systems may operate for a long time. For instance, a Fire Brigade may delay the update of the communication systems embedded in a truck for budget reasons. Such asynchronous evolution issues are critical at the SoS level.

The coupling matrix provides a helpful insight to tackle such issues. Let us illustrate that on the EFS scenario. Table 3 lists the various systems' owners and managers.

Table 3: Systems' owners and providers.

|  | Emergency Operation Command Center (1) | Mobile Head Quarter (2), | Air Officer (3) | Depart-mental Operation Center (4) | Command Helicopter (5) | Coordi-nation plane (6) | Ground firefighter squads (7) | Weather team (8) | Water bomber (9) |
|---|---|---|---|---|---|---|---|---|---|
| Owner | Fire brigade | Fire Brigade | Civil Air Traffic Mana-gement Office | Local Civil Authority | Fire Brigade | Fire Brigade | Fire Brigade | Local Civil Authority | Homeland security agency (nation A) |
| Provider | Industrial company (Civil Security System Inc.) | Industrial company (C3I Ltd.) | Industrial company (ATM Gmbh. | Industrial company (Civil Security System Inc.) | Helicopter manu-facturer (Heli Corp.) | Aircraft manu-facturer (Nice Plane Ltd.) | Manu-facturers (Teleco SA, Truck Gmbh., Incendio Vestiti S.p.A.) | Industrial company (Frog SA) | Aircraft manu-facturer (Shower Plane Ltd.) |

In addition to their variety, it should be noted that the systems themselves are part of different SoS. For instance, each plane manufacturer manages its transmission systems with its suppliers. This concerns the type of data, their semantic, syntax and format.

Moreover, the evolution of the mobile HQ system may be the result of economic constraints. If the mobile HQ system provider updates the system, what are the impacts of this updating upon the interfaces with the other systems? Do these impacts necessarily imply evolution of the other related systems?

The coupling matrix helps answering such questions: if we read it as a process dependency matrix, the presence of many non-diagonal terms emphasizes tight coupling. On the contrary, a sparse matrix means a weak coupling. Thus the coupling matrix visually identifies sets of component systems whose collaboration is both essential and complex for the achievement of the emerging services. One of several possible SoS architecting processes can be defined:
1. Develop the scenarios describing the critical emerging functions.
2. Draw the resulting coupling matrix.
3. Identify the sets of strongly coupled systems (by permuting the systems, so as transform the coupling matrix into a block-diagonal matrix, as described in [MEI02]).
4. Once the critical sets are identified, adopt one or more of the following policies :
   - At least, each owner/manager of a component belonging to a critical set should be very cautious when designing a new version, and verify that interoperability is still ensured;
   - If it is possible, merge the management of the systems to ensure consistency of evolutions;
   - If it is possible, change the perimeter of the systems by merging them, from a technical point of view.

When looking at this example, we observe a weak coupling between the ground firefighter squads and the other systems. This is due to autonomy needs and ad hoc and interactive interaction, because effectors are human and the environment evolves dynamically. On the contrary, the ground firefighter squads are linked to the mobile head quarter, so, if this mobile HQ evolves, that implies that ground firefighter squads evolve too. It is necessary to assess the impacts of the evolution of the former to the later. In this case, who is responsible for the evolutions of these systems? Who pays for them?

If the customer systems do not evolve at the same level, there is an asynchronous evolution of the SoS and an increased risk as to SoS capabilities. Lack of compliance means lack of interoperability, whence loss of emergent capabilities of the SoS. The asynchronous evolution issue is very important since owners and managers of the various component systems are different, with strong aims and constraints upon the systems. To deal with this problem, an independent organization, such as a state or an agency, may impose a globally planned and orchestrated evolution, resolving such asynchronous evolution issues. The european project OASIS (Open Advanced System for dIsaster and emergency management, http://www.oasis-fp6.org) is an example to resolve such an issue. The objective of OASIS is to define and develop an Information Technology framework based on an open and flexible architecture and using standards, existing or proposed by OASIS. That will be the basis of a European Disaster and Emergency Management system.

Typically, when two systems are strongly coupled and exhibit weak coupling with the remaining systems, it could be appropriate to look at both systems globally and have a common organization level responsible for managing them. Such a question has its importance when addressing the communication infrastructure in a service-oriented architecture: who holds responsibility for this key asset of the SoS and manages its configuration, in adequacy with all other evolutions? A straightforward solution would be to have a contractor assuming integration responsibility for each subset of strongly coupled systems. Whether this can be applied to all problems is another question: on the one hand, it can be an advantage for Defense & Security SoS, or highly regulated SoS such as the air transportation and air traffic management SoS, but on the other hand it is an obvious barrier to spontaneous emergence of new behaviors.

## Coupling matrix, asynchronous evolution and failure mode definition

The previous section hinted at how the coupling matrix could exhibit the impacts of asynchronous evolution, including emergent risks, as asynchronous evolution may degrade the whole SoS performance and safety. Indeed, as Levenson et al. [LEV06] write: "often, degradation of the control structure involves asynchronous evolution, where one part of a system changes without the related necessary changes in other parts. Changes to subsystems may be carefully designed, but consideration of their effects on other parts of the system, including the control aspects, may be neglected or inadequate". Asynchronous evolution may trigger a failure of the interface between the related systems and imply by cascade effect a failure of the whole SoS. In this case, the new desired capabilities are not available, and worse, some critical component systems might have a failure that would not have occurred under stand-alone conditions.

A formal verification procedure based on formal techniques cited before can provide a static failure analysis of the SoS. The dependence coupling matrix is a useful representation for this, as it enhances readability of the compositional behavior. However it does not give any answer concerning dynamic failure analysis, since the dynamic environment of a behavior during execution cannot be captured unfortunately by static descriptions.

The only interesting answer, easily provided by our approach, is the search for a priori responsibilities between owners and managers in case of an identified failure. Indeed the incriminated service (either provided by a system, or arising as an emerging SoS service by combination of existing services) yields a set of potential responsible actors and resources, identified by reading the cells of the matrix. This raises the issue of responsibility transfers, which can be partially solved when an LSI (lead system integrator) or another appropriate risk-sharing or risk-assuming entity is designated. Back to our example assuming the SoS emerging services are in place, that allows a "real-time" reassignment of the firefighting means following the weather monitoring and forecast. If the fire spreads, who should carry the responsibility… and the authority to decide actions?

An indication of the "manageability" of a safe SoS could be the number of individual interactions involving two different owners, and therefore requiring the establishment of a contract. This is enlightened by colored cells in the following table.

Table 4: Ownership and SoS service paths.

| 1<br>Fire brigade | [1.1], [1.3]<br>**Internal** |  | [1.2], [1.4]<br>**Contract** |  |  |  |  |
|---|---|---|---|---|---|---|---|
| [2.1]<br>**Internal** | 2<br>Fire brigade |  | [2.3]<br>**Contract** |  | [2.2], [2.4]<br>**Contract** |  | [2.4]<br>**Contract** |
|  | [3.1]<br>**Contract** | 3<br>Civil ATM Office |  |  | [3.1]<br>**Contract** |  | [3.2]<br>**Contract** |
|  | [4.4]<br>**Contract** |  | 4<br>Local Civil Authority |  | [4.2], [4.3]<br>**Contract** |  |  |
|  | [5.1], [5.2]<br>**Internal** |  |  | 5<br>Fire brigade |  |  |  |
|  |  | [6.3]<br>**Contract** |  |  | 6<br>Fire brigade |  | [6.1], [6.2]<br>**Contract** |
|  | [7.1]<br>**Internal** |  |  |  |  | 7<br>Fire brigade |  |
| [8.2]<br>**Contract** |  | [8.2]<br>**Contract** | [8.2]<br>**Internal** |  |  |  | 8<br>Local Civil Authority | [8.2]<br>**Contract** |
|  | [9.1]<br>**Contract** | [9.2]<br>**Contract** |  |  |  |  |  | 9<br>Homeland security agency |

This rather obvious remark should be correlated with the impact analysis mentioned before,

which relies on the connectivity degree of the various services: integrators carrying the responsibility should be designated a priori for the services with the highest connectivity index.

# Further extensions: towards a dynamical view of SoS management

Up to now we have dealt with static representations, isolating in time a specific view of the SoS. When addressing configuration management, a temporal analysis is more appropriate, especially for SoS which raise major agility issues: in that case each component (including the environment) is subject to radical changes in time that impact the whole SoS. This can occur at the (logical and physical) architecture level, at the mission level (evolving context of use and change of operating rules), at the organization level (vanishing and arising actors and/or contractors), etc. The instantaneous versions of the coupling matrix and its various enrichments can be assembled into a time-indexed bundle, and this new representation provides novel ways to tackle complex issues such as business strategic or acquisition issues: if system integrators are defined, their responsibility perimeter can be easily correlated to the evolution of the coupling within the SoS. For instance, the coloring mentioned in the previous section that helped seeing the connected business partners, defines in this higher-dimensional representation various colored subspaces (obtained by stacking the individual colored regions), which can be analyzed graphically very easily in terms of intertwining or connectivity.

Although this may seem a little far-fetched, it is a research direction worth to pursue, especially when one remembers that sustainability of complex systems is currently critical, as it concentrates the major part of the budgetary resources of the life-cycle for systems, and that we lack tools to manage such issues.

# Biography

Frédéric Autran, born in 1961, earned the engineer diploma granted by Ecole Centrale de Paris in 1984. After 5 years in CASE tool development, he acted as a consultant for the DGA and contributed to the building of a semantic interoperability framework among French C3I. He then entered EADS and is in charge of system engineering methods and tools for the Defense and Communication Systems business unit. He set up and leads the System of Systems and Complex Systems working group of AFIS since 2005.

Jean-Philippe Auzelle has 15 years experience in manufacturing education. He was an academic referent for French education (from 2000 to 2005) and was in charge to implement best practices in computer-supported cooperative work for this institution. He earned a Master degree in Mechanical Engineering from the National Polytechnic Institute of Lorraine (France) in 2005. He joined the Research Center of Automatic Control in Nancy since 2005 to prepare his PhD, with Pr. Morel, in the "Product Driven Systems" team-project. His research interests include "Ambient Manufacturing Systems" to model, to design and to evaluate the interoperability between enterprise software. He is currently involved in the design framework of new product driven automation architectures.

Denise CATTAN is in charge of the Maturity Model and Indicators working group at AFIS and is scientific adviser at SPIRULA, providing consultancy to improve systems engineering processes. Denise worked for THALES to deploy Systems Engineering Method and to assess maturity of practices based upon the CMMI model (she was member of the CMMI Integrated Product Team, involved as author in the CMMI development). She had been working for twenty years in the development of Short Range Air Defense Weapon Systems. Prior to that, she was in charge of simulation and software development. Denise is graduated in Nuclear Physics from EPF School in 1966.

Dominique Luzeaux graduated from École Polytechnique (1987), École Nationale Supérieure des Techniques Avancées (1989), PhD from University Paris XI (1991). He held the position of an invited postdoctoral fellow at the University of Berkeley in 1991-1992. In 2001 he completed his professor thesis at the University Paris XI. He was Head of the Complex System Engineering Department within the Ministry of Defense from 2002 to 2004, in charge of the R&T programs on Modeling & Simulation, System Engineering and Battle-labs. He is currently director of the

Technical Center for Information Systems. He also teaches robotics and theoretical computer science at graduate level.

Jean-Luc Garnier is the technical coordinator of the THALES architectural advanced studies. His spatiality is engineering and architecting of real-time distributed systems. He has an engineer degree in computer science from INSA (French Institute of Applied Science). From 1984 to 1999, he had been software engineer and expert in consulting companies, mainly in compilers, operating systems and real-time telecoms developments. He joined THALES in 2000 as architect, successively in Integrated Modular Avionics, Electronic Warfare and Network Centric Warfare. He teaches System Architecting for the Master program in Engineering of Complex Industrial Systems (MISIC) of The French Polytechnic School, and the THALES University.

Dr. Frédérique Mayer is associate professor in Information Systems and Systems Engineering at the industrial systems and innovation engineering school (ENSGSI) of the National Polytechnic Institute of Lorraine (INPL). She is researcher in the innovative process research team of INPL and she has a strong experience in system oriented modeling process and enterprise modeling, with special research interests in enterprise systems interoperability and systems approaches. Frédérique Mayer is currently the chairperson of the technical committee 9.2 on "Social Impact of Automation" of IFAC (International Federation of Automatic Control). She is member of the French association of systems engineering and takes part in the working group on system of systems and complex systems.

Marc Peyrichon has been working as a military naval systems of systems architect in DCNS SA (France) since 2005. Prior to this, he had been working in the field of system engineering in various areas such as ship design (1988), enterprise information systems (1991), military radiocommunications and electronic warfare (1996). He was also responsible for revamping the engineering process and relevant PLM tools for DCNS SA design office (2000). He intervenes in UBS university (Vannes-Lorient) as an expert in system engineering.

Jean-René Ruault, earned social psychological master by EHESS (Paris). From 1991 to 2004, he had been software engineer and HCI expert. Since 2004, he has work as systems engineer and standardization manager of the system of systems (SoS) skill center at DGA. Its job deals with human factor engineering and network centricity issues too. He has been AFIS System of Systems / Complex Systems working group co-chairman since 2006, and BNAE Systems Design working group chairman. Christophe KOLSKI, Eric BRANGIER, and he were co-chairmen of the ERGO IA 2006 conference. He has published articles dealing with HCI, and systems engineering.